\documentclass[fleqn,usenatbib]{mnras}

\usepackage{newtxtext,newtxmath}

\usepackage[T1]{fontenc}

\DeclareRobustCommand{\VAN}[3]{#2}
\let\VANthebibliography\thebibliography
\def\thebibliography{\DeclareRobustCommand{\VAN}[3]{##3}\VANthebibliography}
\DeclareMathOperator{\sech}{sech}

\usepackage{graphicx}	
\usepackage{amsmath}	

\newcommand{\kms}{km\,s$^{-1}$}

\title[The centroid and group speed of fast wave trains]{The centroid speed as a characteristic of the group speed of solar coronal fast magnetoacoustic wave trains}

\author[D.~Y. Kolotkov et al.]{
Dmitrii Y. Kolotkov$^{1, 2} \thanks{E-mail: D.Kolotkov.1@warwick.ac.uk},$
Valery M. Nakariakov$^{1} \thanks{E-mail: V.M.Nakariakov@warwick.ac.uk}$,
Maximilien Cloesen$^{1,3}$ 
\\
$^{1}$Centre for Fusion, Space and Astrophysics, Physics Department, University of Warwick, Coventry CV4 7AL, UK\\
$^{2}$ Engineering Research Institute \lq\lq Ventspils International Radio Astronomy Centre (VIRAC)\rq\rq, Ventspils University of Applied Sciences, Ventspils, LV-3601, Latvia\\
$^{3}$ School of Mathematics, Monash University, Victoria 3800, Australia
}

\date{Accepted XXX. Received YYY; in original form ZZZ}

\pubyear{2023}

\begin{document}
\label{firstpage}
\pagerange{\pageref{firstpage}--\pageref{lastpage}}
\maketitle

\begin{abstract}
The highly-filamented nature of the coronal plasma significantly influences dynamic processes in the corona such as magnetohydrodynamic waves and oscillations. Fast magnetoacoustic waves, guided by coronal plasma non-uniformities, exhibit strong geometric dispersion, forming quasi-periodic fast-propagating (QFP) wave trains. QFP  wave trains are observed in extreme-ultraviolet imaging data and indirectly in microwaves and low-frequency radio, aiding in understanding the magnetic connectivity, energy, and mass transport in the corona. However, measuring the field-aligned group speed of QFP wave trains, as a key parameter for seismological analysis, is challenging due to strong dispersion and associated rapid evolution of the wave train envelope. We demonstrate that the group speed of QFP wave trains formed in plane low-$\beta$ coronal plasma non-uniformities can be assessed through the propagation of the wave train's effective centre of mass, referred to as the wave train's centroid speed. This centroid speed, as a potential observable, is shown empirically to correspond to the group speed of the most energetic Fourier harmonic in the wave train. The centroid speed is found to be almost insensitive to the waveguide density contrast with the ambient corona, and to vary with the steepness of the transverse density profile. 
The discrepancy between the centroid speed as the group speed measure and the phase speed at the corresponding wavelength is shown to reach 70\%, which is crucial for the energy flux estimation and interpretation of observations.
\end{abstract}




\section{Introduction}

The elastic and compressible plasma of the corona of the Sun is known to support the propagation of various types of magnetohydrodynamic (MHD) waves \citep[e.g.,][]{2020ARA&A..58..441N}.
Coronal MHD waves are a subject to intensive ongoing studies in the context of the coronal heating problem \citep[e.g.,][]{2020SSRv..216..140V}, and also as natural plasma diagnostic tools \citep[e.g.,][]{2005RSPTA.363.2743D}. 
Properties of the waves are highly affected by perpendicular non-uniformity of equilibrium plasma parameters, such as the density and temperature \citep[e.g.,][]{1982SoPh...76..239E, 1983SoPh...88..179E}. Thus, the field-aligned filamentation of the solar corona in a form of various plasma loops, plumes, etc., plays a decisive role in the MHD wave processes in that plasma environment \citep[e.g.,][]{2016SSRv..200...75N}. In particular, coronal plasma non-uniformities act as fast magnetoacoustic waveguides \citep[e.g.,][]{1988A&A...192..343E}. The fast waves which, in a uniform medium, are perpendicular or oblique, become parallel to the magnetic field because of the reflection or refraction on the non-uniformity of the fast speed.

Guided fast waves are well resolved in the corona in a form of quasi-periodic fast-propagating (QFP) disturbances of the extreme-ultraviolet (EUV) emission intensity \citep[e.g.,][]{2011ApJ...736L..13L, 2012ApJ...753...53S, 2013SoPh..288..585S, 2012ApJ...753...52L, 2017ApJ...851...41Q, 2019ApJ...873...22S, 2019ApJ...871L...2M, 2020ApJ...889..139M, 2020ChSBu..65.3909S}. Typically, QFP waves resemble a train of \lq\lq ripples\rq\rq, emanating from an epicentre in an active region. The waves travel along coronal loops or fan structures, extended in the apparent direction of the magnetic field.  Often, the perpendicular size of the observed wave front increases with the distance from the driver, with the typical expansion angle of about a few tens of degrees \citep[e.g.,][]{2022SoPh..297...20S}. Typical relative amplitudes of the EUV intensity perturbations are 1\%--8\%. Typical projected phase speeds are higher than several hundred \kms. The latter property clearly distinguishes the QFP waves from another propagating wave phenomenon observed in the solar corona, the slow magnetoacoustic waves which propagate at the speed lower than a few hundred \kms\ \citep[e.g.,][]{2006RSPTA.364..461D, 2021SSRv..217...76B}. In some cases, QFP waves and slow waves are detected to propagate simultaneously along the same coronal plasma structure \citep[e.g.,][]{2015A&A...581A..78Z}. The oscillation periods range from several tens to several hundred seconds. A QFP with a much shorter oscillation period, of about 6~s, and the phase speed about 2100~\kms, travelling along an active region coronal loop, was observed by \citet{2002MNRAS.336..747W} in the white light intensity during a solar eclipse.  
Usually, QFP waves show several consecutively propagating wave fronts, and last for a few oscillation cycles only, which is another difference with slow waves which typically last for several tens of cycles at least. However, in some cases, a QFP wave train has more than ten wave fronts \citep{2014A&A...569A..12N}. In addition, there is another kind of QFP waves detected in EUV, which are seen to propagate apparently across the local magnetic field \citep{2022SoPh..297...20S}.  These so-called broad QFP waves have a relative amplitude of up to 30\%, and a larger angular extent of about 90$^\circ$--360$^\circ$.

Sometimes, quasi-periodic pulsations (QPP) often observed in light curves of solar flares detected in various observational bands \citep[see][for a recent comprehensive review]{2021SSRv..217...66Z}, have time signatures typical for QFP waves. For example, \citet{2017ApJ...844..149K} observed a QPP pattern with oscillation periods about 100~s in the microwave, decimetric and soft X-ray emissions. Almost simultaneously, a QFP wave with the instantaneous period decreasing from 240~s to 120~s, and the phase speed about 1000~\kms\ appeared in the EUV. This finding supports the association of QPP with QFP waves in cases when the latter is not detected \citep[e.g.,][]{2009ApJ...697L.108M, 2011SoPh..273..393M, 2018ApJ...861...33K, 2019ApJ...872...71Y}.

QFP waves could be driven by either periodic or impulsive energy releases, see \citet{2011ApJ...740L..33O, 2018ApJ...860...54O} and, for example, \citet{2004MNRAS.349..705N, 2013A&A...554A.144Y, 2014A&A...569A..12N}, respectively. Theoretical modelling of the latter mechanism demonstrated the formation of a quasi-periodic wave train from an impulsive initial perturbation \citep[e.g.,][]{1993SoPh..143...89M, 1993SoPh..144..101M, 1993SoPh..144..255M, 2005SSRv..121..115N, 2014A&A...569A..12N}. This effect is an intrinsic feature of the guided fast magnetoacoustic wave propagation, caused by the wave dispersion, i.e., the dependence of the phase and group speeds on the oscillation period or the wavelength along the waveguide \citep[e.g.,][]{1984ApJ...279..857R, 2014ApJ...789...48O, 2020SSRv..216..136L}. The wave dispersion is caused by the presence of the characteristic spatial scale in the system, the perpendicular width of the waveguide. This dispersion mechanism appears in ideal MHD, and is not caused by the Hall effect or electron inertia. Typically, fast wave trains have an asymmetric envelope, and show the variation of the instantaneous oscillation period. Both effects cease with narrowing the initial spectrum of the driver \citep{2005SSRv..121..115N}.  

Characteristic signatures of the dispersively formed fast wave train are determined by the perpendicular profile of the fast speed \citep[e.g.,][]{1995SoPh..159..399N, 2017ApJ...836....1Y, 2018ApJ...855...53L}. For smoother profiles, the wavelet power spectrum of the fast wave train has a characteristic \lq\lq tadpole\rq\rq\ shape in both slab and cylindrical geometries, see, e.g., \citep{2004MNRAS.349..705N, 2022MNRAS.515.4055G} and \citep{2015ApJ...814..135S}, respectively, and in a slab with a current sheet \citep{2012A&A...537A..46J, 2014ApJ...788...44M}. This feature is consistent with those detected in some observations \citep[e.g.,][]{2004MNRAS.349..705N, 2017ApJ...844..149K}. For steeper profiles, the wavelet spectrum has a characteristic \lq\lq boomerang\rq\rq\ shape \citep{2021MNRAS.505.3505K}, which has also been observationally detected \citep[e.g.,][]{2011SoPh..273..393M}. Essentially, the wavelet signature is determined by the dependence of the group speed of the guided fast wave upon the parallel wavenumber \citep{2004MNRAS.349..705N}. This effect takes place in 2D non-uniformities too, e.g., in magnetic funnels \citep{2013A&A...560A..97P}. 
Fast wave trains driven by a quasi-periodic driver modelled by \citep[e.g.,][]{2011ApJ...740L..33O, 2012ApJ...753...52L} are also consistent with the observed behaviour of QFP waves. A periodically driven wave train does not demonstrate a significant evolution. It is consistent with the dispersive evolution model, as in a narrowband signal, group speeds of spectral harmonics do not differ much from each other. 

The projected phase speed of QFP waves could be readily estimated by measuring the angle of the diagonal ridges in the time-distance map constructed along the wave path in observational data. However, coronal seismology by QFP waves, such as estimating the parameters of the perpendicular non-uniformity of the plasma which is crucial for revealing the nature of coronal loops and the heating mechanism, also requires the estimation of the group speed. Furthermore, the energy flux in the waves is determined by the group speed \citep[e.g.,][]{1995SoPh..161..269L}, which is crucial for assessing the role of QFP waves in coronal heating \citep[][]{2020SSRv..216..140V}. Its estimation by the phase speed \citep[as in, e.g.,][]{2018ApJ...860...54O} relies on the assumption that those speeds have close values. But, previous theoretical modelling demonstrated that group and phase speeds of guided fast waves can significantly differ from each other \citep[e.g.,][]{1995SoPh..159..399N, 2017ApJ...836....1Y, 2018ApJ...855...53L}. In particular, the difference between the phase and group speeds affects the energy flux estimation, as it depends on the value of the speed to the power of three \citep[see, e.g., Eq. (1) of][]{2018ApJ...860...54O}.
Thus, there is a need for a practical recipe for estimating the group speed in imaging data. This procedure is not trivial, as the textbook definition of the group speed as the speed of an envelope of a wave packet is based on the assumption that the wave spectrum is narrow. However, fast magnetoacoustic waves with the parallel wavelengths comparable to the perpendicular width of the waveguide experience high dispersion. Impulsively excited, i.e., broadband fast wave trains have spectral components with group speeds different by a factor of more than two from each other. This means that the wave train envelope experiences rapid evolution and is not symmetric. In particular, the maximum of the perturbation propagates at a speed different from the centre of the wave train and of its leading and trailing edges. 

The aim of this paper is to develop a technique for estimating the group speed of guided fast wave trains in plasma non-uniformities of the solar corona, and to assess its difference with the phase speed that one should expect in the data analysis.
In the study, we adapt the concept of a \lq\lq centrovelocity\rq\rq\ related to the centroid of the pulse in the time and spatial domains and used in, for example, Geophysics for characterising the wave energy transport in dissipative and dispersive media \citep[e.g.,][]{2010Geop...75...37C}. In Sec.~\ref{sec:centroid}, we describe the theoretical model and introduce the concept of a centroid velocity of QFP wave trains in the solar corona. In Sec.~\ref{Sec:centroid-group}, we demonstrate the link between the wave train's centroid speed and the group speed of the most energetic parallel spatial harmonic. Section~\ref{sec:disc} provides a brief summary of the obtained results, discussion and conclusions.

\section{The concept of the centroid speed of guided fast wave trains}
\label{sec:centroid}

In the magnetically dominated coronal plasma, i.e., with the plasma parameter $\beta\to 0$, the dynamics of linear fast magnetoacoustic waves guided by a plasma slab stretched along an equilibrium magnetic field, which represents a coronal plasma non-uniformity, is described by the 2D wave equation,
\begin{equation}\label{eq:2dwave}
    \frac{\partial^2v_x}{\partial t^2}-C_\mathrm{A}^2(x)\left[\frac{\partial^2v_x}{\partial x^2} + \frac{\partial^2v_x}{\partial z^2}\right]=0,
\end{equation}
where $v_x$ stands for the perturbation of the perpendicular plasma velocity, the $z$-axis coincides with the direction of the guiding magnetic field $B_0$, and the $x$-axis represents the direction of the cross-field plasma density enhancement, $\rho_0(x)$. The total pressure balance requires the equilibrium magnetic field to be constant. Hence, the enhancement of the equilibrium plasma density at a certain location across the field results in the local depletion of the Alfv\'en speed, $C_\mathrm{A}(x)=B_0/\sqrt{\mu_0\rho_0(x)}$. Thus, a field-aligned enhancement of the zero-$\beta$ plasma density is a cavity or a waveguide for fast magnetoacoustic waves \citep[e.g.,][]{2016SSRv..200...75N}. We approximate the equilibrium perpendicular density profile $\rho_0(x)$ by the generalised symmetric Epstein function \citep{1995SoPh..159..399N},
\begin{equation}\label{eq:dens_profile}
    \rho_0(x)=(\rho_\mathrm{in}-\rho_\mathrm{ext})\sech^2\left[\left(\frac{x}{w}\right)^p\right] + \rho_\mathrm{ext},
\end{equation}
where $w$ is the characteristic half-width of the slab, $\rho_\mathrm{in}$ and $\rho_\mathrm{ext}$ are the values of the plasma density at $x=0$ and $x\to\infty$ (corresponding to the Alfv\'en speeds $C_\mathrm{A0}$ and $C_\mathrm{A\infty}$, respectively). The parameter $p$ controls the steepness of the perpendicular density profile. In the limit $p\gg 1$, the profile becomes a step-function. The eigenvalue problem for the profile given by Eq.~(\ref{eq:dens_profile}), describing dispersion relations and the perpendicular structure of guided fast magnetoacoustic waves, has exact analytical solutions for $p\to \infty$ \citep{1982SoPh...76..239E} and $p=1$ \citep{1995SoPh..159..399N, 2003A&A...409..325C}. As the profile of $\rho_0(x)$ is symmetric, it is convenient to distinguish between kink and sausage perturbations, with $d\,v_x(x=0)/dx=0$ and $v_x(x=0)=0$, respectively.

The model with the generalised Epstein profile, i.e., with an arbitrary value of $p$, has been proposed by \citet{1995SoPh..159..399N}, and used in a number of studies. However, in contrast to, for example, \citet{2014A&A...567A..24H, 2021MNRAS.505.3505K}, we do not make the Fourier transform along the slab axis in Eq.~(\ref{eq:2dwave}). An initial value problem constituted by the 2D wave equation is solved numerically with the procedure \textit{ND-Solve} in the \textit{Wolfram Mathematica 12} environment, keeping the functional dependence on all three variables $t$, $x$, and $z$. The described approach, on one hand, allows us to study the evolution of the entire broadband perturbation in time and space (not just the evolution of a single Fourier harmonic along the slab axis) with the subsequent fast wave train formation, and, on the other hand, is more computationally effective than modelling fast wave trains in terms of full MHD.

\begin{figure}
    \centering
    \includegraphics[width=\columnwidth]{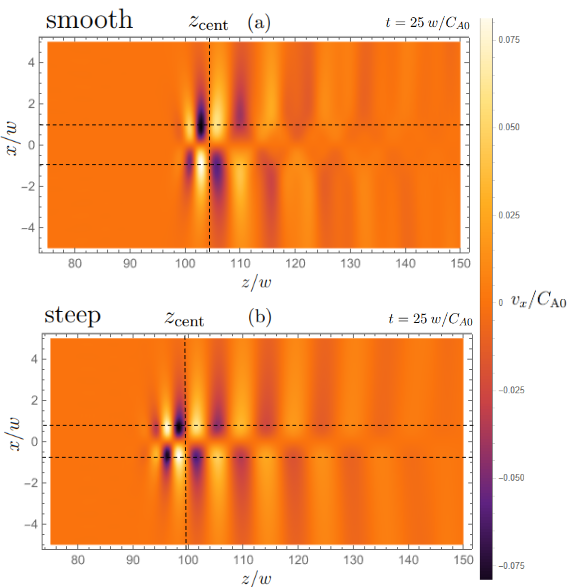}
    \caption{Perturbation of the perpendicular plasma velocity $v_{x}$ at time $t=25\:w/C_\mathrm{A0}$ in an impulsively excited and dispersively evolving fast magnetoacoustic wave train (also referred to as quasi-periodic fast-propagating wave in observations), guided by a field-aligned plasma slab with smooth (a) and steep (b) perpendicular profiles of the plasma density, according to Eq.~(\ref{eq:2dwave}).
    The horizontal black dashed lines correspond to $x=\pm x_{\text{max}}$, the value of $x$ at which $\lvert v_{x} \rvert $ is maximum for all values of $z$; $|x_{\text{max}}|=1.02\,w, 0.79\,w$ in panels (a) and (b), respectively. $z_{\text{cent}}$ denotes the position of the centroid, approximately equal to {$104\,w$} and {$100\,w$} in panels (a) and (b), respectively. The parameter $w$ denotes the half-width of the slab, and $C_\mathrm{A0}$ is the Alfv\'en speed at the axis of the slab.}
    \label{fig:wavetrains2d}
\end{figure}

Snapshots of a fully developed fast wave train of the sausage symmetry, excited by an impulsive driver
\begin{equation}\label{eq:driver}
    v_x(x,z,t=0)=A_0\, x \exp\left[-\left(\frac{x}{d_x}\right)^2\right]\exp\left[-\left(\frac{z-z_0}{d_z}\right)^2\right],
\end{equation}
are shown in Fig.~\ref{fig:wavetrains2d} for a smooth ($p=1$) and steep ($p=5$) density profiles, and the density ratio $\rho_\mathrm{in}/\rho_\mathrm{ext}=10$. In Eq.~(\ref{eq:driver}), $A_0$ is the arbitrary amplitude, the parameters $d_x=w$ and $d_z=\sqrt{2}w$ determine the width of the initial pulse in the $x$ and $z$ directions, respectively. The initial pulse is centered at the axis of the slab ($x=0$) and shifted by $z_0=75\,w$ along the slab axis from the origin. The computational domain extends from $-50\,w$ to $50\,w$ in $x$ and from 0 to $150\,w$ in $z$.

Due to the broadband nature of the driver given by Eq.~(\ref{eq:driver}), multiple Fourier harmonics along the slab axis are excited within a broad range of parallel wavenumbers $k_z$.  Eventually, parallel harmonics with $k_z$ shorter than a certain cut-off value prescribed by the parameters of the perpendicular density profile leak out of the slab. The cut-off wavenumbers are
\begin{align}
        &k_c^\mathrm{smooth}=\frac{1}{w}\sqrt{\frac{2C_{\mathrm{A0}}^{2}}{C_{\mathrm{A\infty}}^{2}-C_{\mathrm{A0}}^{2}}} ,\label{eq:cutoff_smooth}\\
        &k_c^\mathrm{steep}=\frac{\pi}{2w}\sqrt{\frac{C_{\mathrm{A0}}^{2}}{C_{\mathrm{A\infty}}^{2}-C_{\mathrm{A0}}^{2}}}  ,\label{eq:cutoff_steep}
\end{align}
for the Epstein and steep profiles, respectively \citep{1995SoPh..159..399N, 1984ApJ...279..857R}. The remaining harmonics, with $k_z\ge k_c$, get trapped inside the waveguide, propagate along it at different group speeds due to dispersion, and collectively form a quasi-periodic fast wave train guided along the slab. Thus, as such a wave train comprises a broad, continuous range of wavenumbers and corresponding group speeds, a meaningful detection of its speed of propagation in observations remains a challenge. On the other hand, we can see in Fig.~\ref{fig:wavetrains2d} that, for example, wave trains in plasma slabs with smoother density profiles appear globally to propagate faster. Thus, for practical purposes, it is useful to characterise the speed of propagation of such quasi-periodic dispersively evolving fast wave trains from such a global point of view, with the focus put on the dynamics of the entire ensemble rather than the dynamics of individual Fourier components constituting it. As such a global measure of the fast wave train dynamics, we suggest to use the position of its centroid determined as the effective \lq\lq centre of mass\rq\rq\ of the wave train.

\begin{figure*}
    \centering
    \includegraphics[width=\textwidth]{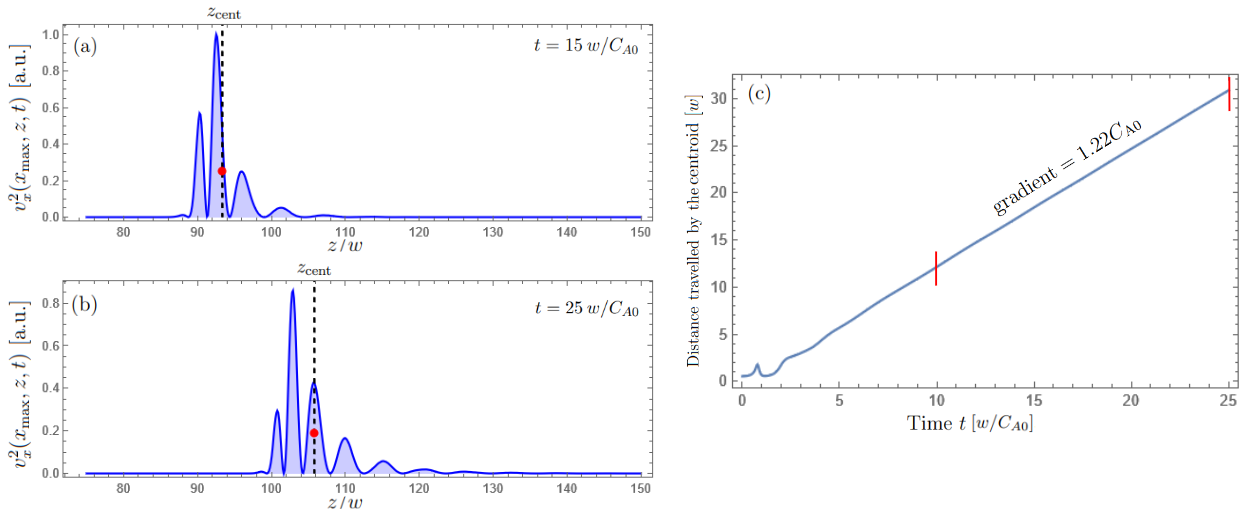}
    \caption{Calculation of the wave train's centroid position and speed. Panels (a) and (b) show $v^2_{x}(x{=}x_{\text{max}},z,t)$ at times $t=15\:w/C_\mathrm{A0}$ and $t=25\:w/C_\mathrm{A0}$. The red dot designates the centroid, with position $z_{\text{cent}}$ (see Sec.~\ref{sec:centroid}, for details). Panel (c) is the time-distance plot for the centroid, the curve is approximated by a linear function in the region delimited by the vertical red bars, with the gradient $v_{\text{cent}}\approx 1.2\:C_\mathrm{A0}$ being the centroid speed.}
    \label{fig:centroid-region}
\end{figure*}

Due to the symmetry of the problem with respect to the axis of the slab, the centroid position in the $x$-direction remains fixed to the axis of the slab, i.e., at $x=0$. Thus, only the wave train's centroid position along the $z$-axis, $z_\mathrm{cent}$ varies with time as the wave train propagates. Hence, we consider the centroid speed $v_\mathrm{cent}=d\,z_\mathrm{cent}/d\,t$. Therefore, finding $v_\mathrm{cent}$ as a characteristic measure of the wave train propagation along the slab axis reduces to the one-dimensional problem of finding $z_\mathrm{cent}(t)$. For this, we fix $x$ to $x_\mathrm{max}$ at which the wave train has the highest amplitude for all $z$ (see Fig.~\ref{fig:wavetrains2d}) and consider $v_x^2(x=x_\mathrm{max},z,t)$ which is an equivalent of the instantaneous kinetic energy density. {The pertrubation of the plasma mass density in the wave train $\rho(x=x_\mathrm{max},z,t)$ can also be used for this analysis, using its linear relationship with $v_x(x,z,t)$ given by Eq.~(7) in \citet{2003A&A...409..325C}, and also \citet{2023MNRAS.520.4147K}. Thus,} for each instant of time, we obtain $z_\mathrm{cent}$ as the weighted centre of the region bounded by the curve $v_x^2(z)$ and the $z$-axis, as illustrated in the left column of Fig.~\ref{fig:centroid-region}. The functions \textit{Polygon} and \textit{Region-Centroid} in \textit{Mathematica 12} are used for this{\footnote{{For each $z_i$ corresponding to the elementary area $a_i$ under the signal of interest, the centroid position $z_\mathrm{cent}$ is calculated as $z_\mathrm{cent} = \Sigma_i a_i z_i / \Sigma_i a_i$. For more details, see \url{https://reference.wolfram.com/language/ref/RegionCentroid.html}.}}.}

The right panel of Fig.~\ref{fig:centroid-region} shows the variation of the wave train's centroid position $z_\mathrm{cent}$ in a waveguide with a smooth density profile with time. The fluctuations seen at $t\lesssim 5w/C_\mathrm{A0}$ are attributed to a combined effect of the initial wave train dispersive formation and leakage of longer-wavelength harmonics. In general, the time scale of such a wave train development could represent another new and potentially interesting observable, but its further discussion is out of the scope of this work. Thus, for estimating $v_\mathrm{cent}$ as the gradient of $z_\mathrm{cent}$ in time, we use the later interval of the obtained $z_\mathrm{cent}(t)$ dependence. In the example illustrated by Fig.~\ref{fig:centroid-region} for a slab with a smooth density profile (the parameter $p=1$ in Eq.~\ref{eq:dens_profile}) and density ratio $\rho_\mathrm{in}/\rho_\mathrm{ext}=10$, the value of $v_\mathrm{cent}$ is found to be about $1.2\, C_\mathrm{A0}$. For a slab with a steep density profile ($p\gg1$) and the same density ratio $\rho_\mathrm{in}/\rho_\mathrm{ext}=10$, the application of the same approach results in $v_\mathrm{cent} \approx 1.0\, C_\mathrm{A0}$.

\section{The centroid and group speed of fast wave trains}
\label{Sec:centroid-group}
In this section, we investigate the physical meaning of the revealed centroid speed $v_\mathrm{cent}$ through its relationship with the group speed $v_\mathrm{g}$ of individual Fourier harmonics constituting the wave train.

\begin{figure*}
    \centering
    \includegraphics[width=0.49\textwidth]{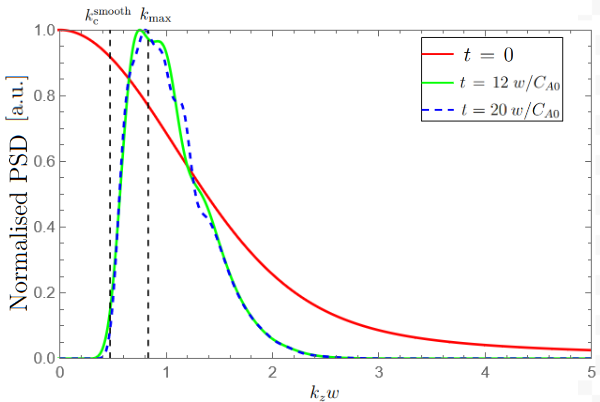}
    \includegraphics[width=0.49\textwidth]{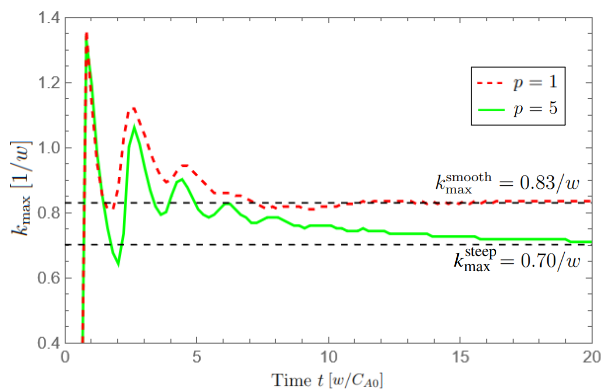}
    \caption{Determining the wavenumber $k_{\text{max}}$ of the most powerful parallel spatial harmonic present in the wave train. Left: normalised power spectral density (PSD) of the wave train for the Epstein transverse density profile at three instances of time; $k^{\text{smooth}}_{c}\approx 0.5/w$ denotes the cut-off wavenumber given by Eqs.~(\ref{eq:cutoff_smooth})--(\ref{eq:cutoff_steep}), and $k_{\text{max}} \approx 0.8/w$ is the wavenumber of the most powerful harmonic at $t \to \infty$. Right: evolution of $k_{\text{max}}$ with time for the Epstein (red, $p=1$) and step-function (green, $p=5$) density profiles. The values $k^{\text{smooth}}_{\text{max}} \approx 0.8/w$ and
    $k^{\text{steep}}_{\text{max}} \approx 0.7/w$ correspond to the value of $k_{\text{max}}$ at $t \to \infty$ for the Epstein and steep density profiles, respectively.}
    \label{fig:fourier_kmax}
\end{figure*}

In order to investigate the behaviour of different Fourier harmonics in the wave train over time, we make the Fourier transform of $v_x(x=x_\mathrm{max},z,t)$ with respect to $z$ (see Fig.~\ref{fig:fourier_kmax}, left panel). For $t=0$, the Fourier power spectrum is of a Gaussian shape which is consistent with the form of the initial broadband perturbation given by Eq.~(\ref{eq:driver}). For later times, the Fourier power density in the longer-wavelength part of the spectrum (for $k_z < k_c$) decreases to almost zero due to the leakage of those harmonics into the external medium. Among the remaining Fourier harmonics with $k_z \ge k_c$, we consider the most powerful one and refer to its wavenumber as $k_\mathrm{max}$. In the initial phase of the wave train development, the value of $k_\mathrm{max}$ varies (see Fig.~\ref{fig:fourier_kmax}, right panel). However, after some transition period, $k_\mathrm{max}$ converges to a constant value which is found to be about $0.8/w$ and $0.7/w$ for the waveguides with the Epstein and step-function density profile, respectively.

\begin{figure*}
    \centering
    \includegraphics[width=\columnwidth]{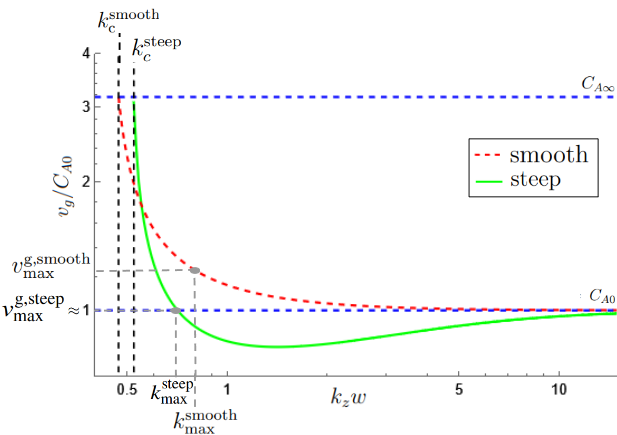}
    \includegraphics[width=\columnwidth]{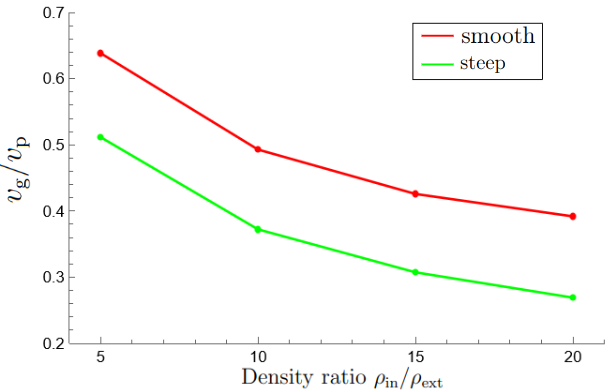}
    \caption{Left: Finding the group speeds $v^{\text{g,smooth}}_{\text{max}}$ and $v^{\text{g,steep}}_{\text{max}}$ of the most powerful harmonic in a fast magnetoacoustic wave train in plasma slabs with the Epstein and step-function profiles. The red dashed and green solid lines show the dependencies of the group speed $v_\mathrm{g}$ on the parallel wavenumber $k_{z}$, obtained with Eqs.~(\ref{eq:disp_smooth})--(\ref{eq:disp_steep}). The values $k^{\text{smooth}}_{\text{max}}=0.8/w$ and $k^{\text{steep}}_{\text{max}}=0.7/w$, obtained in Fig.~\ref{fig:fourier_kmax}, correspond to the group speeds $v^{\text{g,smooth}}_\text{max}\approx 1.2\, C_\mathrm{A0}$ and $v^{\text{g,steep}}_\text{max}\approx C_\mathrm{A0}$, respectively.
    Right: Dependence of the group speed to phase speed ratio $v_{\text{g}}/v_{\text{p}}$, evaluated at $k_{\text{max}}$, upon the waveguide's density contrast, for the Epstein (red) and step-function (green) density profiles.}
    \label{fig:dispersion}
\end{figure*}

We now determine the group speed $v_\mathrm{g}=d\,\omega/d\,k_z$ of the most powerful Fourier harmonic with the parallel wavenumber $k_z=k_\mathrm{max}$ in a fully developed wave train by solving the dispersion relation for a smooth density profile with $p=1$, i.e., the Epstein profile, 
\begin{equation}\label{eq:disp_smooth}
    \frac{|k_z|w}{C_\mathrm{A0}^2}\left(V_\mathrm{p}^2-C_\mathrm{A0}^2\right)=\frac{3}{C_\mathrm{A\infty}}\sqrt{C_\mathrm{A\infty}^2-V_\mathrm{p}^2}+\frac{2}{|k_z|w},
\end{equation}
where $V_\mathrm{p}=\omega/k_z$ is the phase speed  \citep{2003A&A...409..325C}. Likewise, we determine the group speed of fast waves guided by a slab with a steep density profile, corresponding to $p\to \infty$,
\begin{equation}\label{eq:disp_steep}
\tan(wq)=-\frac{q}{m},
\end{equation}
where
\begin{align}
    &q^2 = \left(\frac{\omega}{C_\mathrm{A0}}\right)^2 - k_z^2~~
    \mathrm{and}~~
    m^2 = k_z^2 - \left(\frac{\omega}{C_\mathrm{A\infty}}\right)^2,\nonumber
\end{align}
\citep{1982SoPh...76..239E}. It is interesting that in the Epstein slab case the dispersion relation can be solved analytically, as a solution of a bi-quadratic equation, while in the step function case the roots of the dispersion relation should be determined numerically. 
The obtained dispersion curves are shown in Fig.~\ref{fig:dispersion} for the density contrast $\rho_\mathrm{in}/\rho_\mathrm{ext}=10$. As one can see, the resulting group speeds $v^\mathrm{max}_\mathrm{g}$ of the most powerful Fourier harmonics with $k_z=k_\mathrm{max}$ are found to be about $1.2\,C_\mathrm{A0}$ and $1.0\, C_\mathrm{A0}$ for the Epstein and step-function density profiles, respectively, which approximately coincide with the values of the centroid speed $v_\mathrm{cent}$ detected in Sec.~\ref{sec:centroid}. In other words, the centroid speed $v_\mathrm{cent}$ of fast magnetoacoustic wave trains, revealed in this work as a potentially observable parameter, could be interpreted as the group speed of the most powerful Fourier harmonic in the wave train propagating along the waveguide.
In addition, we use Eq.~(\ref{eq:disp_smooth})--(\ref{eq:disp_steep}) to illustrate the ratio between the guided fast wave group speed and phase speed, both taken at the wavenumber $k_z=k_\mathrm{max}$ as most pronounced in observations (Fig.~\ref{fig:dispersion}, right panel). As one can see, the ratio $V_\mathrm{g}/V_\mathrm{p}$ decreases with the density contrast and steepness of the waveguide. For example, for the waveguide with a steep density profile and density contrast $\rho_\mathrm{in}/\rho_\mathrm{ext}=20$, it drops below 0.3.

\begin{figure}
    \centering
    \includegraphics[width=\columnwidth]{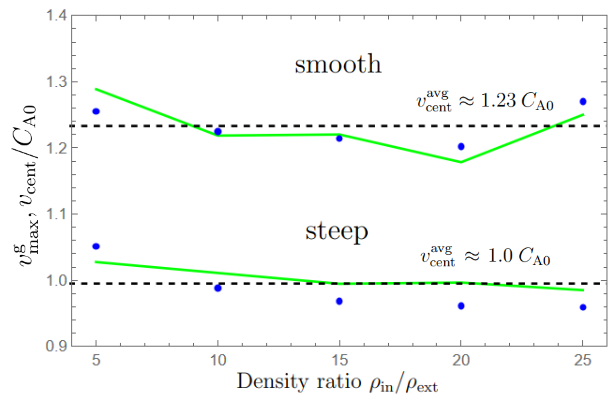}
    \caption{Comparison of the group speed $v^{\text{g}}_{\text{max}}$ of the most powerful harmonic (green line) and the corresponding centroid speed $v_{\text{cent}}$ (blue dots) for varying density ratios, shown for the Epstein and step-function  density profiles. The average centroid speeds are given by $v^{\text{avg}}_\text{cent} \approx 1.2\:C_{\text{A0}}$ and $v^{\text{avg}}_\text{cent}\approx \:C_{\text{A0}}$ for smooth and steep density profiles, respectively.} 
    \label{fig:centroid-group}
\end{figure}


We now investigate whether the relationship between the wave train's centroid speed $v_\mathrm{cent}$ and the group speed of the wave train's most powerful harmonic $v^\mathrm{max}_\mathrm{g}$, revealed empirically for the  density contrast $\rho_\mathrm{in}/\rho_\mathrm{ext}=10$, persists for other density contrasts. For this, we perform the analysis described in Sec.~\ref{sec:centroid} and \ref{Sec:centroid-group} for a broad range of $\rho_\mathrm{in}/\rho_\mathrm{ext}$, from 5 to 25, typical for the Sun's corona. As shown by Fig.~\ref{fig:centroid-group}, for all density contrasts considered, the wave train's centroid speed $v_\mathrm{cent}$ appears to be about the group speed of the most energetic Fourier harmonic in the wave train, $v^\mathrm{max}_\mathrm{g}$. Indeed, the obtained values of $v_\mathrm{cent}$ and $v^\mathrm{max}_\mathrm{g}$, according to Fig.~\ref{fig:centroid-group}, exhibit minor fluctuations in the vicinity of $1.2\,C_\mathrm{A0}$ (for the Epstein profile) and $C_\mathrm{A0}$ (for the step-function profile), and differ from each other by less than 3.5\% which is intrinsically below the expected observational uncertainties of the instrumental or noisy origin. The revealed tendency holds true for the waveguides with smooth and steep perpendicular density profiles.

\section{Discussion and Conclusions}
\label{sec:disc}

We studied the group speed of broadband fast magnetoacoustic wave trains guided by coronal plasma non-uniformities of plane geometry in the low-$\beta$ regime, seen as the phenomenon of quasi-periodic fast-propagating (QFP) waves in multi-band observations. As such impulsively excited QFP wave trains are subject to strong dispersion caused by the waveguiding effect, the main aim of this work was to identify a potential observable which can be used as a characteristic measure of the wave train's group speed. The main findings of this study can be summarised as follows:

\begin{itemize}
    \item Despite each individual Fourier harmonic constituting the wave train propagates at its own speed due to the effect of dispersion, the wave train in the waveguide with a smooth density profile is found to propagate globally faster than the wave train in the waveguide with a steep density profile.

    \item As such a global characteristic of the wave train's dispersive dynamics in the waveguide, we identified the position of its effective centre of mass and referred to its speed of propagation along the waveguide axis as the wave train's centroid speed. This centroid speed of a fully developed wave train is found to be about the internal Alfv\'en speed, $C_\mathrm{A0}$ for waveguides with steep (step-function) density profiles. In consistence with the above finding based on the visual inspection, the wave train's centroid speed for a smooth (Epstein-function) density profile is found to be about 20\% higher, i.e., about $1.2\, C_\mathrm{A0}$. Interestingly, the wave train's centroid position varies non-monotonically with quasi-periodic fluctuations at the very beginning of the wave train's evolution, when it is subject to a combined effect of dispersion and leakage of longer-wavelength harmonics. The time scale of this settlement process may represent another potentially interesting observable which would require a dedicated follow-up study.

    \item The revealed centroid speed of such dispersively evolving QFP wave trains is shown to coincide approximately with the group speed of the most energetic Fourier harmonic in the wave train for both smooth and steep transverse density profiles. This result has important implications for a more meaningful estimation of the energy flux carried by the wave train. Indeed, in contrast to previous works which are based on the assumption that the observed transverse propagating waves in coronal plasma structures are just weakly dispersive and hence the group speed can be approximated by the observed phase speed \citep[e.g.,][]{2014ApJ...795...18V, 2018ApJ...860...54O}, observations of the centroid speed would allow for obtaining a lower bound estimate of the actual energy flux. Our study shows that the group to phase speed ratio of the most energetic Fourier harmonic in the wave train (expected to be most pronounced in observations) can readily be lower than 0.3. The latter would result in more than an order of magnitude discrepancy if one uses the phase speed instead of the group speed in the energy flux estimation.
    
    \item  The group speed of the most energetic Fourier harmonic in the wave train (characterised by the wave train's centroid speed) shows almost no dependence on the density contrast inside and outside the waveguide, which allows for reducing the number of free parameters in seismological analysis \citep[cf.][who proposed the use of the minimum in the group speed dip which is sensitive to both the density profile steepness and contrast]{2016ApJ...833...51Y, 2021MNRAS.505.3505K}. In other words, observations of the wave train's centroid speed may help us with assessing the local internal Alfv\'en speed, $C_\mathrm{A0}$ within 20\% uncertainty (as it is found to vary from $C_\mathrm{A0}$ to $1.2\, C_\mathrm{A0}$, according to our analysis). Or, if $C_\mathrm{A0}$ can be obtained from independent observations, its comparison with the centroid speed can be used for discriminating between steep and smooth transverse density profiles of the wave-hosting plasma structure.
    
\end{itemize}

The performed analysis is based on a plasma slab model, i.e., plane geometry, which is typical, for example, for streamers and/or current sheets. Thus, in addition to the application of the revealed centroid speed to observations, the next natural step would be to generalise this study for cylindrical geometry to account for broader range of coronal plasma structures. However, as properties of coronal QFP wave trains in plane and cylindrical geometries were previously shown to be qualitatively similar \citep[e.g.,][]{2018ApJ...855...53L}, we expect only minor quantitative differences.

\section*{Acknowledgements}

We acknowledge funding from the STFC consolidated grant ST/X000915/1 (D.Y.K.) and Latvian Council of Science Project No. lzp2022/1-0017 (D.Y.K. and V.M.N.). This research was also partly supported by the International Space Science Institute (ISSI) in Bern, through ISSI International Team project \#546 \lq\lq Magnetohydrodynamic Surface Waves at Earth’s Magnetosphere and Beyond\rq\rq\ (V.M.N.). The authors also thank Mr Chengyi Sun, who participated in this research together with M.C. as final-year (MPhys) undergraduate students in the Physics Department of the University of Warwick.

\section*{Data Availability}
The data underlying this article are available in the article and in the references therein.



\bibliographystyle{mnras}

\begin{thebibliography}{}
\makeatletter
\relax
\def\mn@urlcharsother{\let\do\@makeother \do\$\do\&\do\#\do\^\do\_\do\%\do\~}
\def\mn@doi{\begingroup\mn@urlcharsother \@ifnextchar [ {\mn@doi@}
  {\mn@doi@[]}}
\def\mn@doi@[#1]#2{\def\@tempa{#1}\ifx\@tempa\@empty \href
  {http://dx.doi.org/#2} {doi:#2}\else \href {http://dx.doi.org/#2} {#1}\fi
  \endgroup}
\def\mn@eprint#1#2{\mn@eprint@#1:#2::\@nil}
\def\mn@eprint@arXiv#1{\href {http://arxiv.org/abs/#1} {{\tt arXiv:#1}}}
\def\mn@eprint@dblp#1{\href {http://dblp.uni-trier.de/rec/bibtex/#1.xml}
  {dblp:#1}}
\def\mn@eprint@#1:#2:#3:#4\@nil{\def\@tempa {#1}\def\@tempb {#2}\def\@tempc
  {#3}\ifx \@tempc \@empty \let \@tempc \@tempb \let \@tempb \@tempa \fi \ifx
  \@tempb \@empty \def\@tempb {arXiv}\fi \@ifundefined
  {mn@eprint@\@tempb}{\@tempb:\@tempc}{\expandafter \expandafter \csname
  mn@eprint@\@tempb\endcsname \expandafter{\@tempc}}}

\bibitem[\protect\citeauthoryear{{Banerjee} et~al.,}{{Banerjee}
  et~al.}{2021}]{2021SSRv..217...76B}
{Banerjee} D.,  et~al., 2021, \mn@doi [\ssr] {10.1007/s11214-021-00849-0},
  \href {https://ui.adsabs.harvard.edu/abs/2021SSRv..217...76B} {217, 76}

\bibitem[\protect\citeauthoryear{{Carcione}, {Gei}  \& {Treitel}}{{Carcione}
  et~al.}{2010}]{2010Geop...75...37C}
{Carcione} J.~M.,  {Gei} D.,   {Treitel} S.,  2010, \mn@doi [Geophysics]
  {10.1190/1.3346064}, \href
  {https://ui.adsabs.harvard.edu/abs/2010Geop...75...37C} {75, T37}

\bibitem[\protect\citeauthoryear{{Cooper}, {Nakariakov}  \&
  {Williams}}{{Cooper} et~al.}{2003}]{2003A&A...409..325C}
{Cooper} F.~C.,  {Nakariakov} V.~M.,   {Williams} D.~R.,  2003, \mn@doi [\aap]
  {10.1051/0004-6361:20031071}, \href
  {https://ui.adsabs.harvard.edu/abs/2003A&A...409..325C} {409, 325}

\bibitem[\protect\citeauthoryear{{De Moortel}}{{De
  Moortel}}{2005}]{2005RSPTA.363.2743D}
{De Moortel} I.,  2005, \mn@doi [Philosophical Transactions of the Royal
  Society of London Series A] {10.1098/rsta.2005.1665}, \href
  {https://ui.adsabs.harvard.edu/abs/2005RSPTA.363.2743D} {363, 2743}

\bibitem[\protect\citeauthoryear{{De Moortel}}{{De
  Moortel}}{2006}]{2006RSPTA.364..461D}
{De Moortel} I.,  2006, \mn@doi [Philosophical Transactions of the Royal
  Society of London Series A] {10.1098/rsta.2005.1710}, \href
  {https://ui.adsabs.harvard.edu/abs/2006RSPTA.364..461D} {364, 461}

\bibitem[\protect\citeauthoryear{{Edwin} \& {Roberts}}{{Edwin} \&
  {Roberts}}{1982}]{1982SoPh...76..239E}
{Edwin} P.~M.,  {Roberts} B.,  1982, \mn@doi [\solphys] {10.1007/BF00170986},
  \href {https://ui.adsabs.harvard.edu/abs/1982SoPh...76..239E} {76, 239}

\bibitem[\protect\citeauthoryear{{Edwin} \& {Roberts}}{{Edwin} \&
  {Roberts}}{1983}]{1983SoPh...88..179E}
{Edwin} P.~M.,  {Roberts} B.,  1983, \mn@doi [\solphys] {10.1007/BF00196186},
  \href {https://ui.adsabs.harvard.edu/abs/1983SoPh...88..179E} {88, 179}

\bibitem[\protect\citeauthoryear{{Edwin} \& {Roberts}}{{Edwin} \&
  {Roberts}}{1988}]{1988A&A...192..343E}
{Edwin} P.~M.,  {Roberts} B.,  1988, \aap, \href
  {https://ui.adsabs.harvard.edu/abs/1988A&A...192..343E} {192, 343}

\bibitem[\protect\citeauthoryear{{Guo}, {Li}, {Van Doorsselaere}  \&
  {Shi}}{{Guo} et~al.}{2022}]{2022MNRAS.515.4055G}
{Guo} M.,  {Li} B.,  {Van Doorsselaere} T.,   {Shi} M.,  2022, \mn@doi [\mnras]
  {10.1093/mnras/stac2006}, \href
  {https://ui.adsabs.harvard.edu/abs/2022MNRAS.515.4055G} {515, 4055}

\bibitem[\protect\citeauthoryear{{Hornsey}, {Nakariakov}  \&
  {Fludra}}{{Hornsey} et~al.}{2014}]{2014A&A...567A..24H}
{Hornsey} C.,  {Nakariakov} V.~M.,   {Fludra} A.,  2014, \mn@doi [\aap]
  {10.1051/0004-6361/201423524}, \href
  {https://ui.adsabs.harvard.edu/abs/2014A&A...567A..24H} {567, A24}

\bibitem[\protect\citeauthoryear{{Jel{\'\i}nek} \&
  {Karlick{\'y}}}{{Jel{\'\i}nek} \& {Karlick{\'y}}}{2012}]{2012A&A...537A..46J}
{Jel{\'\i}nek} P.,  {Karlick{\'y}} M.,  2012, \mn@doi [\aap]
  {10.1051/0004-6361/201117883}, \href
  {https://ui.adsabs.harvard.edu/abs/2012A&A...537A..46J} {537, A46}

\bibitem[\protect\citeauthoryear{{Kaltman} \& {Kupriyanova}}{{Kaltman} \&
  {Kupriyanova}}{2023}]{2023MNRAS.520.4147K}
{Kaltman} T.~I.,  {Kupriyanova} E.~G.,  2023, \mn@doi [\mnras]
  {10.1093/mnras/stad421}, \href
  {https://ui.adsabs.harvard.edu/abs/2023MNRAS.520.4147K} {520, 4147}

\bibitem[\protect\citeauthoryear{{Kolotkov}, {Nakariakov}  \&
  {Kontar}}{{Kolotkov} et~al.}{2018}]{2018ApJ...861...33K}
{Kolotkov} D.~Y.,  {Nakariakov} V.~M.,   {Kontar} E.~P.,  2018, \mn@doi [\apj]
  {10.3847/1538-4357/aac77e}, \href
  {https://ui.adsabs.harvard.edu/abs/2018ApJ...861...33K} {861, 33}

\bibitem[\protect\citeauthoryear{{Kolotkov}, {Nakariakov}, {Moss}  \&
  {Shellard}}{{Kolotkov} et~al.}{2021}]{2021MNRAS.505.3505K}
{Kolotkov} D.~Y.,  {Nakariakov} V.~M.,  {Moss} G.,   {Shellard} P.,  2021,
  \mn@doi [\mnras] {10.1093/mnras/stab1587}, \href
  {https://ui.adsabs.harvard.edu/abs/2021MNRAS.505.3505K} {505, 3505}

\bibitem[\protect\citeauthoryear{{Kumar}, {Nakariakov}  \& {Cho}}{{Kumar}
  et~al.}{2017}]{2017ApJ...844..149K}
{Kumar} P.,  {Nakariakov} V.~M.,   {Cho} K.-S.,  2017, \mn@doi [\apj]
  {10.3847/1538-4357/aa7d53}, \href
  {https://ui.adsabs.harvard.edu/abs/2017ApJ...844..149K} {844, 149}

\bibitem[\protect\citeauthoryear{{Laing} \& {Edwin}}{{Laing} \&
  {Edwin}}{1995}]{1995SoPh..161..269L}
{Laing} G.~B.,  {Edwin} P.~M.,  1995, \mn@doi [\solphys] {10.1007/BF00732071},
  \href {https://ui.adsabs.harvard.edu/abs/1995SoPh..161..269L} {161, 269}

\bibitem[\protect\citeauthoryear{{Li}, {Guo}, {Yu}  \& {Chen}}{{Li}
  et~al.}{2018}]{2018ApJ...855...53L}
{Li} B.,  {Guo} M.-Z.,  {Yu} H.,   {Chen} S.-X.,  2018, \mn@doi [\apj]
  {10.3847/1538-4357/aaaf19}, \href
  {https://ui.adsabs.harvard.edu/abs/2018ApJ...855...53L} {855, 53}

\bibitem[\protect\citeauthoryear{{Li}, {Antolin}, {Guo}, {Kuznetsov}, {Pascoe},
  {Van Doorsselaere}  \& {Vasheghani Farahani}}{{Li}
  et~al.}{2020}]{2020SSRv..216..136L}
{Li} B.,  {Antolin} P.,  {Guo} M.~Z.,  {Kuznetsov} A.~A.,  {Pascoe} D.~J.,
  {Van Doorsselaere} T.,   {Vasheghani Farahani} S.,  2020, \mn@doi [\ssr]
  {10.1007/s11214-020-00761-z}, \href
  {https://ui.adsabs.harvard.edu/abs/2020SSRv..216..136L} {216, 136}

\bibitem[\protect\citeauthoryear{{Liu}, {Title}, {Zhao}, {Ofman}, {Schrijver},
  {Aschwanden}, {De Pontieu}  \& {Tarbell}}{{Liu}
  et~al.}{2011}]{2011ApJ...736L..13L}
{Liu} W.,  {Title} A.~M.,  {Zhao} J.,  {Ofman} L.,  {Schrijver} C.~J.,
  {Aschwanden} M.~J.,  {De Pontieu} B.,   {Tarbell} T.~D.,  2011, \mn@doi
  [\apjl] {10.1088/2041-8205/736/1/L13}, \href
  {https://ui.adsabs.harvard.edu/abs/2011ApJ...736L..13L} {736, L13}

\bibitem[\protect\citeauthoryear{{Liu}, {Ofman}, {Nitta}, {Aschwanden},
  {Schrijver}, {Title}  \& {Tarbell}}{{Liu} et~al.}{2012}]{2012ApJ...753...52L}
{Liu} W.,  {Ofman} L.,  {Nitta} N.~V.,  {Aschwanden} M.~J.,  {Schrijver} C.~J.,
   {Title} A.~M.,   {Tarbell} T.~D.,  2012, \mn@doi [\apj]
  {10.1088/0004-637X/753/1/52}, \href
  {https://ui.adsabs.harvard.edu/abs/2012ApJ...753...52L} {753, 52}

\bibitem[\protect\citeauthoryear{{M{\'e}sz{\'a}rosov{\'a}}, {Karlick{\'y}},
  {Ryb{\'a}k}  \& {Ji{\v{r}}i{\v{c}}ka}}{{M{\'e}sz{\'a}rosov{\'a}}
  et~al.}{2009}]{2009ApJ...697L.108M}
{M{\'e}sz{\'a}rosov{\'a}} H.,  {Karlick{\'y}} M.,  {Ryb{\'a}k} J.,
  {Ji{\v{r}}i{\v{c}}ka} K.,  2009, \mn@doi [\apjl]
  {10.1088/0004-637X/697/2/L108}, \href
  {https://ui.adsabs.harvard.edu/abs/2009ApJ...697L.108M} {697, L108}

\bibitem[\protect\citeauthoryear{{M{\'e}sz{\'a}rosov{\'a}}, {Karlick{\'y}}  \&
  {Ryb{\'a}k}}{{M{\'e}sz{\'a}rosov{\'a}} et~al.}{2011}]{2011SoPh..273..393M}
{M{\'e}sz{\'a}rosov{\'a}} H.,  {Karlick{\'y}} M.,   {Ryb{\'a}k} J.,  2011,
  \mn@doi [\solphys] {10.1007/s11207-011-9794-6}, \href
  {https://ui.adsabs.harvard.edu/abs/2011SoPh..273..393M} {273, 393}

\bibitem[\protect\citeauthoryear{{M{\'e}sz{\'a}rosov{\'a}}, {Karlick{\'y}},
  {Jel{\'\i}nek}  \& {Ryb{\'a}k}}{{M{\'e}sz{\'a}rosov{\'a}}
  et~al.}{2014}]{2014ApJ...788...44M}
{M{\'e}sz{\'a}rosov{\'a}} H.,  {Karlick{\'y}} M.,  {Jel{\'\i}nek} P.,
  {Ryb{\'a}k} J.,  2014, \mn@doi [\apj] {10.1088/0004-637X/788/1/44}, \href
  {https://ui.adsabs.harvard.edu/abs/2014ApJ...788...44M} {788, 44}

\bibitem[\protect\citeauthoryear{{Miao}, {Liu}, {Shen}, {Li}, {Abidin},
  {Elmhamdi}  \& {Kordi}}{{Miao} et~al.}{2019}]{2019ApJ...871L...2M}
{Miao} Y.~H.,  {Liu} Y.,  {Shen} Y.~D.,  {Li} H.~B.,  {Abidin} Z.~Z.,
  {Elmhamdi} A.,   {Kordi} A.~S.,  2019, \mn@doi [\apjl]
  {10.3847/2041-8213/aafaf9}, \href
  {https://ui.adsabs.harvard.edu/abs/2019ApJ...871L...2M} {871, L2}

\bibitem[\protect\citeauthoryear{{Miao} et~al.,}{{Miao}
  et~al.}{2020}]{2020ApJ...889..139M}
{Miao} Y.,  et~al., 2020, \mn@doi [\apj] {10.3847/1538-4357/ab655f}, \href
  {https://ui.adsabs.harvard.edu/abs/2020ApJ...889..139M} {889, 139}

\bibitem[\protect\citeauthoryear{{Murawski} \& {Roberts}}{{Murawski} \&
  {Roberts}}{1993a}]{1993SoPh..143...89M}
{Murawski} K.,  {Roberts} B.,  1993a, \mn@doi [\solphys] {10.1007/BF00619098},
  \href {https://ui.adsabs.harvard.edu/abs/1993SoPh..143...89M} {143, 89}

\bibitem[\protect\citeauthoryear{{Murawski} \& {Roberts}}{{Murawski} \&
  {Roberts}}{1993b}]{1993SoPh..144..101M}
{Murawski} K.,  {Roberts} B.,  1993b, \mn@doi [\solphys] {10.1007/BF00667986},
  \href {https://ui.adsabs.harvard.edu/abs/1993SoPh..144..101M} {144, 101}

\bibitem[\protect\citeauthoryear{{Murawski} \& {Roberts}}{{Murawski} \&
  {Roberts}}{1993c}]{1993SoPh..144..255M}
{Murawski} K.,  {Roberts} B.,  1993c, \mn@doi [\solphys] {10.1007/BF00627592},
  \href {https://ui.adsabs.harvard.edu/abs/1993SoPh..144..255M} {144, 255}

\bibitem[\protect\citeauthoryear{{Nakariakov} \& {Kolotkov}}{{Nakariakov} \&
  {Kolotkov}}{2020}]{2020ARA&A..58..441N}
{Nakariakov} V.~M.,  {Kolotkov} D.~Y.,  2020, \mn@doi [\araa]
  {10.1146/annurev-astro-032320-042940}, \href
  {https://ui.adsabs.harvard.edu/abs/2020ARA&A..58..441N} {58, 441}

\bibitem[\protect\citeauthoryear{{Nakariakov} \& {Roberts}}{{Nakariakov} \&
  {Roberts}}{1995}]{1995SoPh..159..399N}
{Nakariakov} V.~M.,  {Roberts} B.,  1995, \mn@doi [\solphys]
  {10.1007/BF00686541}, \href
  {https://ui.adsabs.harvard.edu/abs/1995SoPh..159..399N} {159, 399}

\bibitem[\protect\citeauthoryear{{Nakariakov}, {Arber}, {Ault}, {Katsiyannis},
  {Williams}  \& {Keenan}}{{Nakariakov} et~al.}{2004}]{2004MNRAS.349..705N}
{Nakariakov} V.~M.,  {Arber} T.~D.,  {Ault} C.~E.,  {Katsiyannis} A.~C.,
  {Williams} D.~R.,   {Keenan} F.~P.,  2004, \mn@doi [\mnras]
  {10.1111/j.1365-2966.2004.07537.x}, \href
  {https://ui.adsabs.harvard.edu/abs/2004MNRAS.349..705N} {349, 705}

\bibitem[\protect\citeauthoryear{{Nakariakov}, {Pascoe}  \&
  {Arber}}{{Nakariakov} et~al.}{2005}]{2005SSRv..121..115N}
{Nakariakov} V.~M.,  {Pascoe} D.~J.,   {Arber} T.~D.,  2005, \mn@doi [\ssr]
  {10.1007/s11214-006-4718-8}, \href
  {https://ui.adsabs.harvard.edu/abs/2005SSRv..121..115N} {121, 115}

\bibitem[\protect\citeauthoryear{{Nakariakov} et~al.,}{{Nakariakov}
  et~al.}{2016}]{2016SSRv..200...75N}
{Nakariakov} V.~M.,  et~al., 2016, \mn@doi [\ssr] {10.1007/s11214-015-0233-0},
  \href {https://ui.adsabs.harvard.edu/abs/2016SSRv..200...75N} {200, 75}

\bibitem[\protect\citeauthoryear{{Nistic{\`o}}, {Pascoe}  \&
  {Nakariakov}}{{Nistic{\`o}} et~al.}{2014}]{2014A&A...569A..12N}
{Nistic{\`o}} G.,  {Pascoe} D.~J.,   {Nakariakov} V.~M.,  2014, \mn@doi [\aap]
  {10.1051/0004-6361/201423763}, \href
  {https://ui.adsabs.harvard.edu/abs/2014A&A...569A..12N} {569, A12}

\bibitem[\protect\citeauthoryear{{Ofman} \& {Liu}}{{Ofman} \&
  {Liu}}{2018}]{2018ApJ...860...54O}
{Ofman} L.,  {Liu} W.,  2018, \mn@doi [\apj] {10.3847/1538-4357/aac2e8}, \href
  {https://ui.adsabs.harvard.edu/abs/2018ApJ...860...54O} {860, 54}

\bibitem[\protect\citeauthoryear{{Ofman}, {Liu}, {Title}  \&
  {Aschwanden}}{{Ofman} et~al.}{2011}]{2011ApJ...740L..33O}
{Ofman} L.,  {Liu} W.,  {Title} A.,   {Aschwanden} M.,  2011, \mn@doi [\apjl]
  {10.1088/2041-8205/740/2/L33}, \href
  {https://ui.adsabs.harvard.edu/abs/2011ApJ...740L..33O} {740, L33}

\bibitem[\protect\citeauthoryear{{Oliver}, {Ruderman}  \& {Terradas}}{{Oliver}
  et~al.}{2014}]{2014ApJ...789...48O}
{Oliver} R.,  {Ruderman} M.~S.,   {Terradas} J.,  2014, \mn@doi [\apj]
  {10.1088/0004-637X/789/1/48}, \href
  {https://ui.adsabs.harvard.edu/abs/2014ApJ...789...48O} {789, 48}

\bibitem[\protect\citeauthoryear{{Pascoe}, {Nakariakov}  \&
  {Kupriyanova}}{{Pascoe} et~al.}{2013}]{2013A&A...560A..97P}
{Pascoe} D.~J.,  {Nakariakov} V.~M.,   {Kupriyanova} E.~G.,  2013, \mn@doi
  [\aap] {10.1051/0004-6361/201322678}, \href
  {https://ui.adsabs.harvard.edu/abs/2013A&A...560A..97P} {560, A97}

\bibitem[\protect\citeauthoryear{{Qu}, {Jiang}  \& {Chen}}{{Qu}
  et~al.}{2017}]{2017ApJ...851...41Q}
{Qu} Z.~N.,  {Jiang} L.~Q.,   {Chen} S.~L.,  2017, \mn@doi [\apj]
  {10.3847/1538-4357/aa9beb}, \href
  {https://ui.adsabs.harvard.edu/abs/2017ApJ...851...41Q} {851, 41}

\bibitem[\protect\citeauthoryear{{Roberts}, {Edwin}  \& {Benz}}{{Roberts}
  et~al.}{1984}]{1984ApJ...279..857R}
{Roberts} B.,  {Edwin} P.~M.,   {Benz} A.~O.,  1984, \mn@doi [\apj]
  {10.1086/161956}, \href
  {https://ui.adsabs.harvard.edu/abs/1984ApJ...279..857R} {279, 857}

\bibitem[\protect\citeauthoryear{{Shen} \& {Liu}}{{Shen} \&
  {Liu}}{2012}]{2012ApJ...753...53S}
{Shen} Y.,  {Liu} Y.,  2012, \mn@doi [\apj] {10.1088/0004-637X/753/1/53}, \href
  {https://ui.adsabs.harvard.edu/abs/2012ApJ...753...53S} {753, 53}

\bibitem[\protect\citeauthoryear{{Shen}, {Liu}, {Su}, {Li}, {Zhang}, {Tian},
  {Zhao}  \& {Elmhamdi}}{{Shen} et~al.}{2013}]{2013SoPh..288..585S}
{Shen} Y.~D.,  {Liu} Y.,  {Su} J.~T.,  {Li} H.,  {Zhang} X.~F.,  {Tian} Z.~J.,
  {Zhao} R.~J.,   {Elmhamdi} A.,  2013, \mn@doi [\solphys]
  {10.1007/s11207-013-0395-4}, \href
  {https://ui.adsabs.harvard.edu/abs/2013SoPh..288..585S} {288, 585}

\bibitem[\protect\citeauthoryear{{Shen}, {Chen}, {Liu}, {Shibata}, {Tang}  \&
  {Liu}}{{Shen} et~al.}{2019}]{2019ApJ...873...22S}
{Shen} Y.,  {Chen} P.~F.,  {Liu} Y.~D.,  {Shibata} K.,  {Tang} Z.,   {Liu} Y.,
  2019, \mn@doi [\apj] {10.3847/1538-4357/ab01dd}, \href
  {https://ui.adsabs.harvard.edu/abs/2019ApJ...873...22S} {873, 22}

\bibitem[\protect\citeauthoryear{{Shen}, {Li}, {Chen}, {Zhou}  \& {Liu}}{{Shen}
  et~al.}{2020}]{2020ChSBu..65.3909S}
{Shen} Y.,  {Li} B.,  {Chen} P.,  {Zhou} X.,   {Liu} Y.,  2020, \mn@doi
  [Chinese Science Bulletin] {10.1360/TB-2020-0748}, \href
  {https://ui.adsabs.harvard.edu/abs/2020ChSBu..65.3909S} {65, 3909}

\bibitem[\protect\citeauthoryear{{Shen}, {Zhou}, {Duan}, {Tang}, {Zhou}  \&
  {Tan}}{{Shen} et~al.}{2022}]{2022SoPh..297...20S}
{Shen} Y.,  {Zhou} X.,  {Duan} Y.,  {Tang} Z.,  {Zhou} C.,   {Tan} S.,  2022,
  \mn@doi [\solphys] {10.1007/s11207-022-01953-2}, \href
  {https://ui.adsabs.harvard.edu/abs/2022SoPh..297...20S} {297, 20}

\bibitem[\protect\citeauthoryear{{Shestov}, {Nakariakov}  \& {Kuzin}}{{Shestov}
  et~al.}{2015}]{2015ApJ...814..135S}
{Shestov} S.,  {Nakariakov} V.~M.,   {Kuzin} S.,  2015, \mn@doi [\apj]
  {10.1088/0004-637X/814/2/135}, \href
  {https://ui.adsabs.harvard.edu/abs/2015ApJ...814..135S} {814, 135}

\bibitem[\protect\citeauthoryear{{Van Doorsselaere}, {Gijsen}, {Andries}  \&
  {Verth}}{{Van Doorsselaere} et~al.}{2014}]{2014ApJ...795...18V}
{Van Doorsselaere} T.,  {Gijsen} S.~E.,  {Andries} J.,   {Verth} G.,  2014,
  \mn@doi [\apj] {10.1088/0004-637X/795/1/18}, \href
  {https://ui.adsabs.harvard.edu/abs/2014ApJ...795...18V} {795, 18}

\bibitem[\protect\citeauthoryear{{Van Doorsselaere} et~al.,}{{Van Doorsselaere}
  et~al.}{2020}]{2020SSRv..216..140V}
{Van Doorsselaere} T.,  et~al., 2020, \mn@doi [\ssr]
  {10.1007/s11214-020-00770-y}, \href
  {https://ui.adsabs.harvard.edu/abs/2020SSRv..216..140V} {216, 140}

\bibitem[\protect\citeauthoryear{{Williams}, {Mathioudakis}, {Gallagher},
  {Phillips}, {McAteer}, {Keenan}, {Rudawy}  \& {Katsiyannis}}{{Williams}
  et~al.}{2002}]{2002MNRAS.336..747W}
{Williams} D.~R.,  {Mathioudakis} M.,  {Gallagher} P.~T.,  {Phillips} K.~J.~H.,
   {McAteer} R.~T.~J.,  {Keenan} F.~P.,  {Rudawy} P.,   {Katsiyannis} A.~C.,
  2002, \mn@doi [\mnras] {10.1046/j.1365-8711.2002.05764.x}, \href
  {https://ui.adsabs.harvard.edu/abs/2002MNRAS.336..747W} {336, 747}

\bibitem[\protect\citeauthoryear{{Yu} \& {Chen}}{{Yu} \&
  {Chen}}{2019}]{2019ApJ...872...71Y}
{Yu} S.,  {Chen} B.,  2019, \mn@doi [\apj] {10.3847/1538-4357/aaff6d}, \href
  {https://ui.adsabs.harvard.edu/abs/2019ApJ...872...71Y} {872, 71}

\bibitem[\protect\citeauthoryear{{Yu}, {Li}, {Chen}, {Xiong}  \& {Guo}}{{Yu}
  et~al.}{2016}]{2016ApJ...833...51Y}
{Yu} H.,  {Li} B.,  {Chen} S.-X.,  {Xiong} M.,   {Guo} M.-Z.,  2016, \mn@doi
  [\apj] {10.3847/1538-4357/833/1/51}, \href
  {https://ui.adsabs.harvard.edu/abs/2016ApJ...833...51Y} {833, 51}

\bibitem[\protect\citeauthoryear{{Yu}, {Li}, {Chen}, {Xiong}  \& {Guo}}{{Yu}
  et~al.}{2017}]{2017ApJ...836....1Y}
{Yu} H.,  {Li} B.,  {Chen} S.-X.,  {Xiong} M.,   {Guo} M.-Z.,  2017, \mn@doi
  [\apj] {10.3847/1538-4357/836/1/1}, \href
  {https://ui.adsabs.harvard.edu/abs/2017ApJ...836....1Y} {836, 1}

\bibitem[\protect\citeauthoryear{{Yuan}, {Shen}, {Liu}, {Nakariakov}, {Tan}  \&
  {Huang}}{{Yuan} et~al.}{2013}]{2013A&A...554A.144Y}
{Yuan} D.,  {Shen} Y.,  {Liu} Y.,  {Nakariakov} V.~M.,  {Tan} B.,   {Huang} J.,
   2013, \mn@doi [\aap] {10.1051/0004-6361/201321435}, \href
  {https://ui.adsabs.harvard.edu/abs/2013A&A...554A.144Y} {554, A144}

\bibitem[\protect\citeauthoryear{{Zhang}, {Zhang}, {Wang}  \&
  {Nakariakov}}{{Zhang} et~al.}{2015}]{2015A&A...581A..78Z}
{Zhang} Y.,  {Zhang} J.,  {Wang} J.,   {Nakariakov} V.~M.,  2015, \mn@doi
  [\aap] {10.1051/0004-6361/201525621}, \href
  {https://ui.adsabs.harvard.edu/abs/2015A&A...581A..78Z} {581, A78}

\bibitem[\protect\citeauthoryear{{Zimovets} et~al.,}{{Zimovets}
  et~al.}{2021}]{2021SSRv..217...66Z}
{Zimovets} I.~V.,  et~al., 2021, \mn@doi [\ssr] {10.1007/s11214-021-00840-9},
  \href {https://ui.adsabs.harvard.edu/abs/2021SSRv..217...66Z} {217, 66}

\makeatother
\end{thebibliography}




%
%
%

\bsp	
\label{lastpage}
\end{document}